\documentclass[a4paper]{spie}  
\usepackage[]{graphicx}

\title{Soft gamma-ray optics: new Laue lens design and performance estimates} 

\author{Nicolas M. Barri{\`e}re\supit{a}, 
Lorenzo Natalucci\supit{a}, 
Nikolay Abrosimov\supit{c}, 
Peter von Ballmoos\supit{b},
Pierre Bastie\supit{d},  
Pierre Courtois\supit{e}, 
Michael Jentschel\supit{e}, 
J{\"u}rgen Kn{\"o}dlseder\supit{b}, 
Julien Rousselle\supit{b} 
and 
Pietro Ubertini\supit{a}
\skiplinehalf
\supit{a}INAF - IASF Roma, via Fosso del Cavaliere 100, 00133 Roma, Italy \\
\supit{b}CESR - UMR 5187, 9 av du Colonel Roche, 31028 Toulouse, France \\
\supit{c}IKZ,  Max Born-Str. 2, D-12489 Berlin, Germany \\
\supit{d}LSP - UMR 5588, 140 Av. de la physique, 38402 Saint Martin d'H{\`e}res, France \\
\supit{e}ILL,  6 rue Jules Horowitz, 38042 Grenoble, France
}

\authorinfo{Further author information: (Send correspondence to N.B. or L.N.)\\ N.B.: E-mail: nicolas.barriere\{at\}iasf-roma.inaf.it \\ L.N.: E-mail: lorenzo.natalucci\{at\}iasf-roma.inaf.it}

\begin{document} 
  \maketitle 

\begin{abstract}
Laue lenses are an emerging technology based on diffraction in crystals that allows the concentration of soft gamma rays. This kind of optics that works in the 100 keV - 1.5 MeV band can be used to realize an high-sensitivity and high-angular resolution telescope (in a narrow field of view). This paper reviews the recent progresses that have been done in the development of efficient crystals, in the design study and in the modelisation of the answer of Laue lenses.
Through the example of a new concept of 20 m focal length lens focusing in the 100 keV - 600 keV band, the performance of a telescope based on a Laue lens is presented. This lens uses the most efficient mosaic crystals in each sub-energy range in order to yield the maximum reflectivity. Imaging capabilities are investigated and shows promising results.

\end{abstract}


\section{INTRODUCTION}
\label{sec:intro} 
A Laue lens concentrates gamma-rays using Bragg diffraction in the volume of a large number of crystals arranged in concentric rings and accurately orientated in order to diffract radiation coming from infinity towards a unique focal point (e.g. Ref. \citenum{lund.1992ft, halloin.2005gd}). This principle is applicable from $\sim$ 100 keV up to 1.5 MeV, but with a unique lens it is difficult to cover efficiently a continuous energy band of more than about half an order of magnitude wide.

Thanks to the decoupling between collecting area and sensitive area, a Laue lens produces a dramatic increase of the signal to background ratio. This is the key to achieve the so awaited sensitivity leap of one - two orders of magnitude with respect to past and current instruments (IBIS and SPI onboard INTEGRAL, CompTel onboard CGRO,  BAT onboard SWIFT). Despite a Laue lens is a concentrator (i.e. not a direct imaging system) it allows the reconstruction of images, as it will be shown in this paper. Finally,  combined with a focal plane capable of tracking the various interaction of an event, the telescope becomes sensitive to the polarization:  the lens being fully transparent to polarization \cite{barriere.2008, curado-da-silva.2008xy} it lets the possibility to analyze it in the focal plane. 

Hence a telescope featuring a Laue lens would be perfectly adapted to further our understanding of high-energy processes occurring in a variety of violent events. High sensitivity investigations of point sources such as compact objects, pulsars and active galactic nuclei should bring important insights into the still poorly understood emission mechanisms. Also the link between jet ejection and accretion in black hole and neutron star systems could be clarified by observation of the spectral properties and polarization of the transition state emission.
Although the search for the origin of galactic positrons is not the main driver for Laue lens design, the sensitivity it provides added to its -poor- imaging capabilities could be of great help to solve this mystery. By scanning potential positron emitters such as galactic black hole binary systems a telescope featuring a Laue lens could bring at least much stronger constraints on the population contributing to the galactic positrons and to their diffusion in the interstellar medium.

This paper reports on the current  study and development status of Laue lenses. Sec. \ref{sec:crystals} proposes a review of the latter experimental results we obtained with high-diffraction efficiency crystals, which are the elementary constituent of these lenses. Sec. \ref{sec:design} presents a general study of the influence of the main design parameters on the resulting telescope sensitivity. Then, in Sec. \ref{sec:LL20m} a 20 m focal length lens focusing between 100 and 600 keV is presented. Performance of the telescope are investigated, especially its sensitivity and imaging capabilities.

\section{OPTICS DEVELOPMENT: CRYSTALS} 
\label{sec:crystals}
\begin{figure}[t]
\begin{center}
\includegraphics[width=0.38\textwidth]{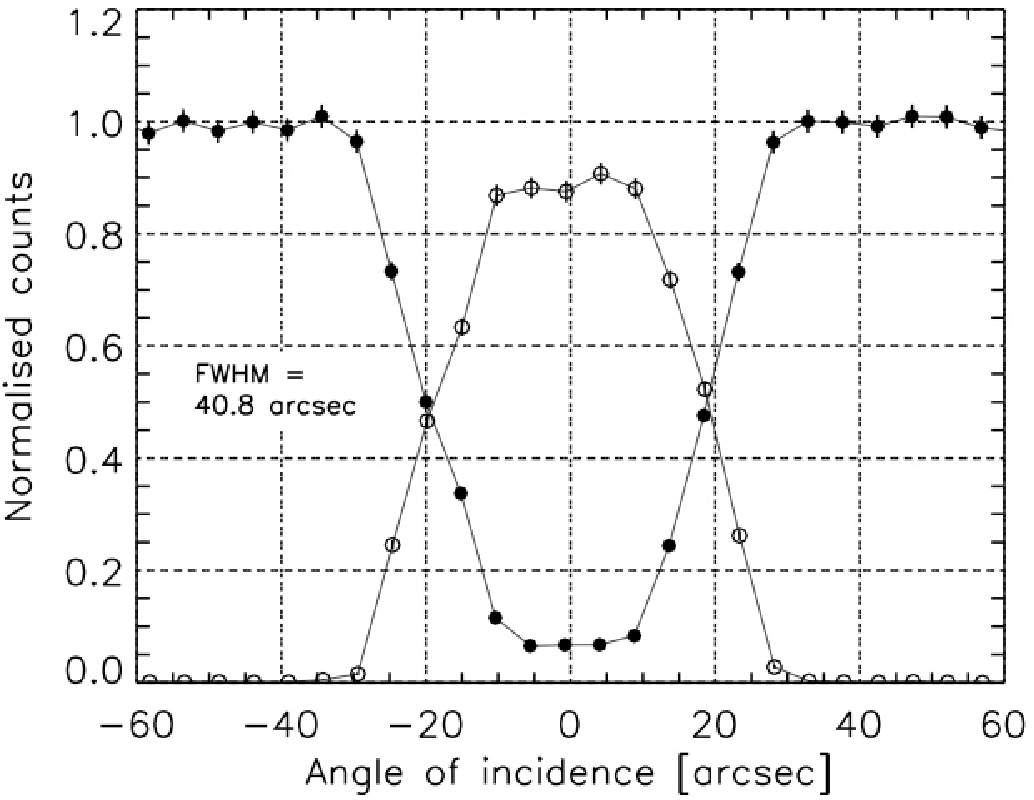}
\hspace{1.2cm}
\includegraphics[width=0.38\textwidth]{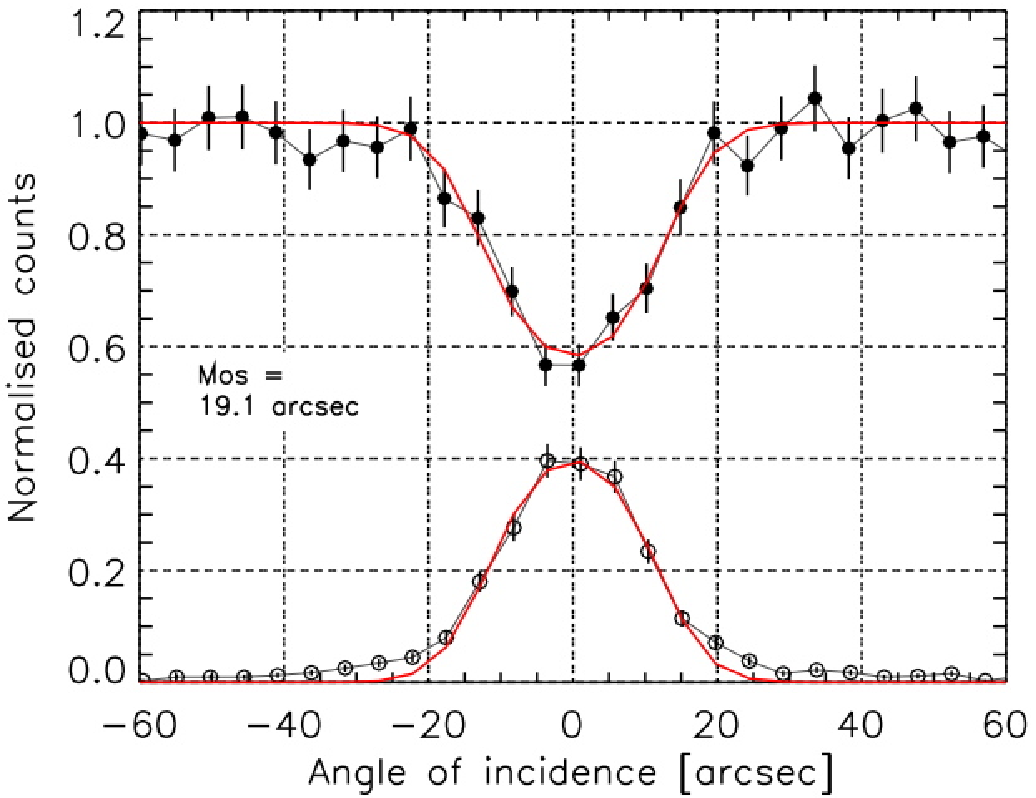}
\caption{\textit{Left:} RC of a 23 mm thick SiGe gradient crystal at 150 keV using the 111 reflection. Diffraction efficiency reach 0.92, which gives a reflectivity of 0.44 once the absorption is included. The FWHM of the RC is 41 arcsec, but as it is visible it is convolved by the divergence of the beam which is responsible of the pyramide-like shape, instead of the expected rectangular shape. \textit{Right:} RC of a 9.7 mm thick Cu mosaic crystal at 500 keV using the 200 reflection. Diffraction efficiency reach 0.40, which gives a reflectivity of 0.20. Mosaicity determined by the fit of Darwin's model (red line) is 19 arcsec.}
\label{fig:RC_cu+SiGe}
\end{center}
\end{figure}

Since the Gamma-Ray Imager (GRI) mission was proposed to the European Space Agency (ESA) in 2007 \cite{knodlseder.2009kx} the endeavor to get efficient and reproducible crystals met some success. First of all the crystals foreseen for the GRI Laue lens, mosaic copper crystals and Si$_{1-x}$Ge$_x$ composition gradient crystals (hereafter referred to as SiGe crystals) were extensively studied first in the framework of a colaboration between the CESR, the ILL and the IKZ\cite{barriere.2007kx}, and then in framework of a dedicated study financed by the ESA. Results are very positive and demonstrate the availability and reproducibility of these two kind of crystals with bandpass of 30 arcsec (private communication). As an example Figure \ref{fig:RC_cu+SiGe} shows two very nice rocking curves\footnote{A rocking curve is obtained by puting a crystal in a monochromatic beam which diffracted intensity is recorded while the crystal is rocked around its Bragg angle. The same measurement is also performed using the transmitted beam in order to get the transmitted intensity outside of the diffraction peak (which gives us the reference value to calculate the diffraction efficiency and the reflectivity).} (RC) recorded during experiment MA-717 on beamline ID15A of the European Synchrotron Radiation Facility (ESRF) on both mosaic Cu and SiGe gradient crystals produced recently respectively at ILL and IKZ.

\begin{figure}[t]
\begin{center}
\includegraphics[width=0.344\textwidth]{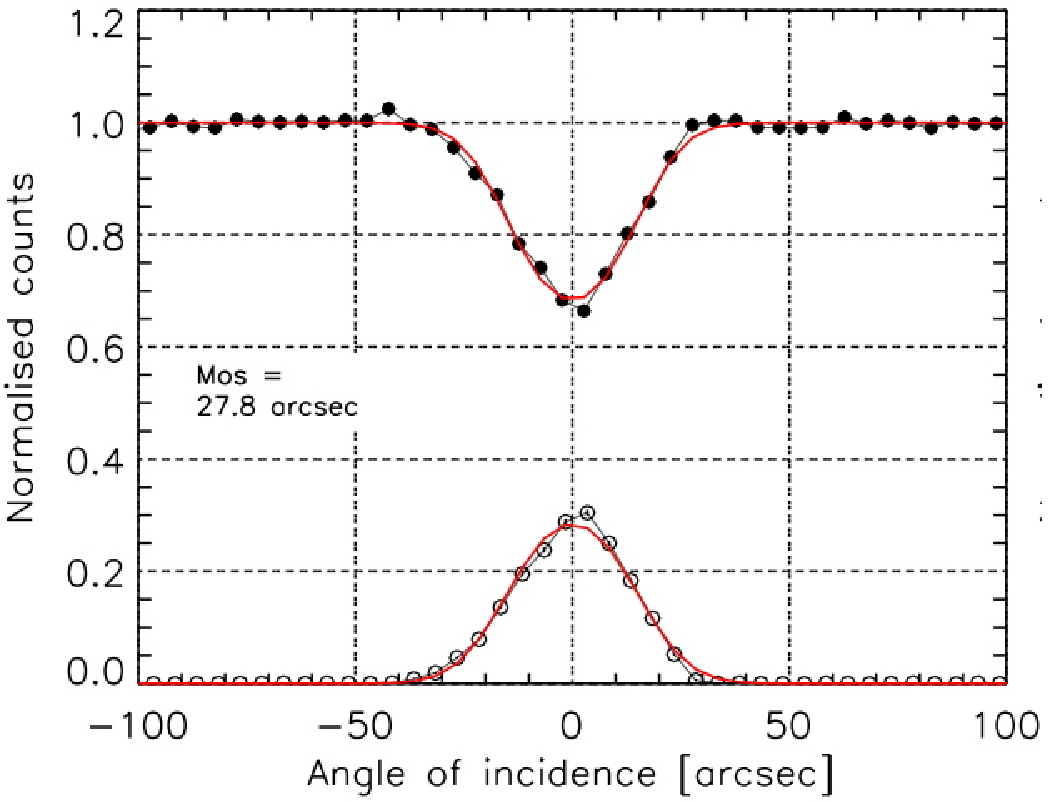}
\includegraphics[width=0.32\textwidth]{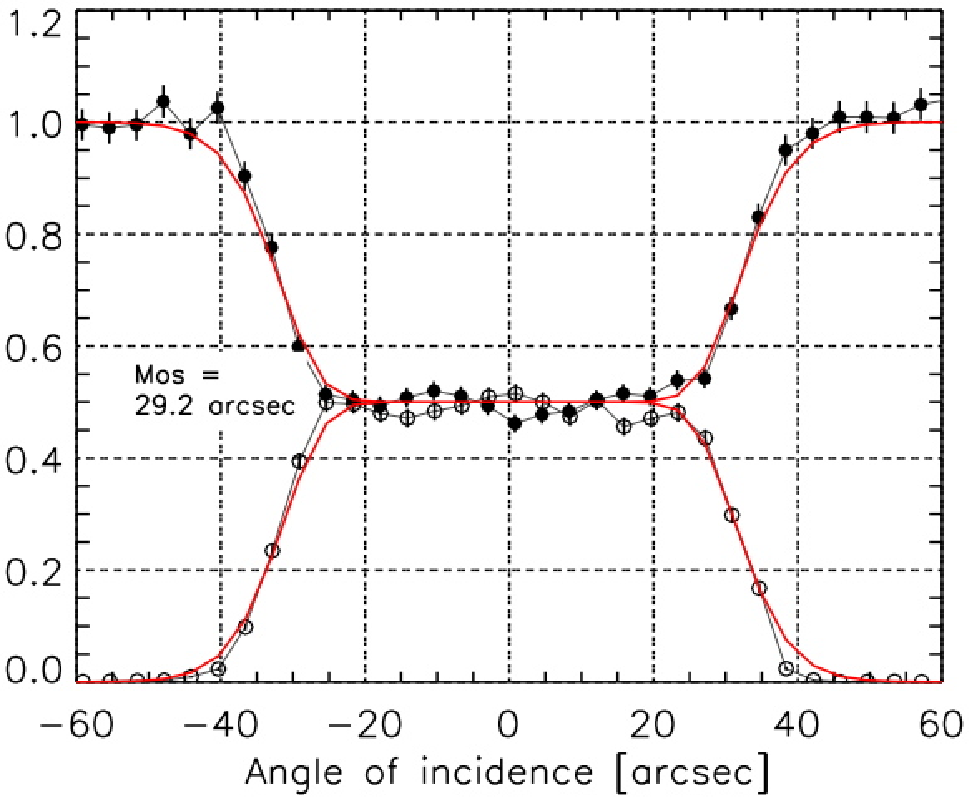}
\includegraphics[width=0.32\textwidth]{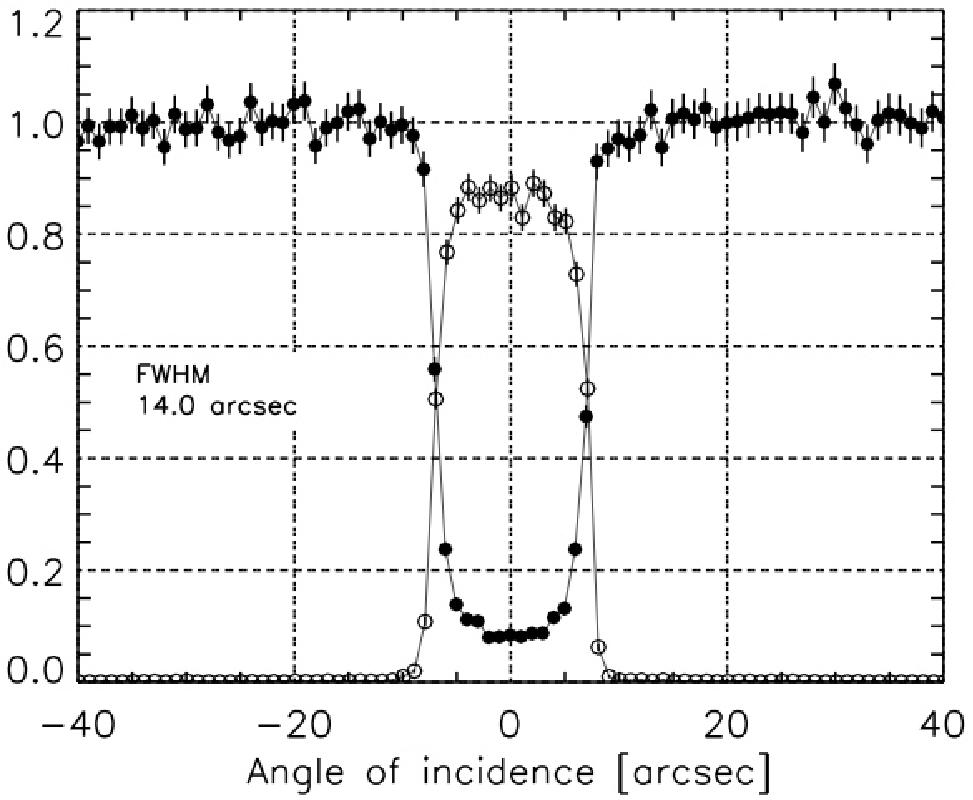}
\caption{\textit{Left:} 
RC of a 10 mm thick Rh crystal at 700 keV using the 220 reflection. Diffraction efficiency reach 0.33, which gives a reflectivity of 0.14 once the absorption is included. This low reflectivity is explained by an over-thickness and by the fact the 220 reflection is not the most efficient. The important here is the low mosaicity of 30 arcsec and the very nice agreement with Darwin's model (red line).
\textit{Centre:} RC of a 25 mm thick GaAs crystal at 200 keV using the 220 reflection. Diffraction efficiency reaches its theoretical maximum of 50\% and the curve is flat-toped because of the too large thickness (with respect to its mosaicity of 29 arcsec and the energy of measurement). Including the absorption, the reflectivity is  6.5 \%. Again, there is a very good agreement with Darwin's model (red line).
\textit{Right:} RC of the first prototype of grooved Si wafer. Diffraction takes place on the (curved) (111) planes across the 10 mm width of the sample. Measurements are done at 150 keV, the diffraction efficiency is 0.9 which gives a reflectivity of 0.65. The curved is rectangular shaped as expected in the case of CDP crystals, with a width of 14 arcsec.}
\label{fig:RC_Rh+GaAs+Sistrie}
\end{center}
\end{figure}

But the scope of the search for crystals did not limit to SiGe and Cu crystals.
Actually the conception of an efficient broad-band lens requires several different crystals in order to utilize each one only at the energy where it is yielding the maximum reflectivity, as shown in Ref. \citenum{barriere.2009sf}. For instance above $\sim$600 keV Cu crystals are not any more the most adapted. Instead gold or silver have a much higher potential (see Ref. \citenum{rousselle.2009} in this session). Moreover, having the choice between several proven crystals to diffract at an energy is an advantage: with various crystals a given energy band can be covered from various radii of the lens, which allows some flexibility in the design process. In our search and development for 'new' crystals we identified three main drivers:
\begin{description}
\item[Crystals that cleave] \textit{i.e.} crystals that split along their crystalline planes. The orientation of crystals on the lens frame would be much facilitated with such crystals since their external faces are representative of their crystalline planes (which is not the case of Cu, SiGe, Ag...). As shown in Sec. \ref{sec:design} the orientation accuracy of crystals is a major issue for the sensitivity and imaging capabilities of the telescope. Unfortunately this feature seems to be associated mainly with two-components crystals\footnote{crystals having more than two components have generally a large crystal cell volume which decreases dramatically their reflectivity, hence only pure elements and two-components crystals are potentially interesting for a Laue lens.} (like CaF$_2$, or NaCl for instance) which can be very efficient at low energy but not any more interesting above $\sim$400 keV.
\item[High-energy crystals.] So far it seems that the best crystals to diffract above 600 keV are Pb, Ir, Pt, Au, W and Ta\cite{barriere.2009sf}. The difficulty comes from the fact that these metals, except from Au and Pb have melting points very elevated, which makes them difficult (and expensive) to grow with the required quality. Pb is the most efficient of this group and it is cheap. But with a hardness of 1.5 Mohs, it is probably too soft to be orientated and glued accurately. The group Pd, Rh and Ag is also interesting but yields a substantially lower reflectivity around 1 MeV than the previously quoted high-Z pure elements.
\item[Crystals having curved diffracting planes.] Up to very recently, SiGe gradient crystals were the only example of crystals having curved diffracting planes (CDP crystals) that we were able to procure up. These crystals are very good at low energy (i.e. between $\sim$100 keV and $\sim$300 keV) but their reflectivity falls down quickly as energy increases due to their low electron density. During our latter ESRF run, we measured a new kind of sample that could open the way toward a new approach to get CDP crystals: the grooving of one face of a wafer creates a convex cylindrical elastic and permanent curvature along the direction of the grooves\cite{bellucci.2003im}. This method, developed at the University of Ferrara - INFN (Ferrara, Italy) has been successfully applied to a 0.9 mm thick Si wafer: this very first sample yielded a reflectivity of 65\% at 150 keV in a bandpass of 14 arcsec (Fig. \ref{fig:RC_Rh+GaAs+Sistrie}). This reflectivity is very close to the one theoretically expected, which shows that the curvature of the diffracting planes is close to the ideal one. The method is very appealing because it allows a very good reproducibility of the curvature and it could be applicable on crystals having an higher electron density (germanium is a good candidate). However some development remains to be done, especially to check that it still works on thicker samples.
\end{description}

Table \ref{tab:cryst_quality} summarizes all the investigations on crystals for a Laue lens that we made over the past 4 years. All these investigations have been conducted at the ESRF (beamline ID-15A) or at ILL (instrument GAMS 4) using beams energies ranging from 150 keV up to 816 keV. Figure \ref{fig:RC_Rh+GaAs+Sistrie} illustrates these results with three RCs: an high-energy crystal (Rh), a low-energy crystal which mass production and quality seems well controlled (GaAs\cite{Ferrari.2008gd}) and our first prototype of grooved Si wafer.

\begin{table}[t]
\begin{minipage}[htdp]{\textwidth}
    \renewcommand{\footnoterule}{} 
\begin{small}
\begin{center}
\begin{tabular}{lp{1.7cm}p{1.9cm}p{3.5cm}p{6cm}}
\hline
Material & Crystal type & Origin of samples & Sample & Qualitative results \\
\hline
\hline
Cu & mosaic & ILL  & Several tens of pieces of various size & Some inhomogeneities in ingots, but mosaicity range is alright (between 15 arcsec and 2 arcmin) and quality is good. \\
Ge$_{1-x}$Si$_x$ & mosaic &  IKZ  & 556 pieces were mounted on the CLAIRE lens prototype\cite{ballmoos.2005ek} & Homogeneity in ingots is not perfect but mosaicity range is alright.\\
Ag & mosaic & MaTecK\footnote{MaTecK GmbH, Im Langenbroich 20, D-52428 Juelich, Germany} & 6 pieces + 1 ingot &  Measured mosaicity ranges between 15 and 60 arcsec. Good quality crystals.\\
Au & mosaic & MaTecK & 3 pieces & Very positive results, see Ref. \citenum{rousselle.2009} \\
Pt &  mosaic & MaTecK & 1 piece & Mosaicity of the order of 1 degree, this sample was not good. \\
Rh &  mosaic & MaTecK & 4 pieces + 1 ingot: 50 mm long, 12 mm diam. & Crystals were not homogeneous but some part had a mosaicity as low as 30 arcsec.  \\
Pb &  mosaic & MaTecK  & 1 ingot: 50 mm long, 12 mm diam. & Crystal twisted, very broad mosaicity (> 1 degree).\\
InP & mosaic & LPCML\footnote{LPCML - Univ Claude Bernard Lyon I, 22 av. G. Berger, 69622 Villeurbanne, France} & 1 wafer, 1 mm thick & Very nice homogeneity and low mosaicity. \\
GaAs & mosaic & IMEM\footnote{IMEM-CNR, Parco Area delle Scienze 37/A, 43010 Parma, Italy} & 5 pieces & Good homogeneity, mosaicity as low as 15 arcsec \\
CZT & mosaic & IMEM & 1 piece & Bad homogeneity, multigrain, mosaicity of $\sim$ 3 arcmin. \\
CaF$_2$ & mosaic & LPCML & 1 piece & Mosaicity ranged between 0.6 and 6 arcmin depending on the reflection considered.  \\
SiGe gradient & CDP & IKZ & Several tens of pieces &  Bandpass is controlled during the growth, quality is excellent.\\
Si grooved & CDP & Univ. Ferrara - INFN\footnote{University of Ferrara - INFN, Via Saragat 1/C, 44100 Ferrara, Italy} & 1 piece & Very high reflectivity measured for the very first try, very encouraging. \\
\hline
\end{tabular}
\end{center}
\label{tab:cryst_quality}
\end{small}
\end{minipage}
\vspace{0.1cm}
\caption{Qualitative overview of experimental results obtained with crystals potentially interesting for a Laue lens.}
\end{table}

\section{GENERAL DESIGN CONSIDERATIONS}
\label{sec:design}

Independently of the choice of the crystals, the performance of a Laue lens focusing in a given energy band is defined by three main parameters:
\begin{itemize}
\item the focal distance,
\item the size of the crystals (considering they have a square-shaped cross section),
\item the mosaicity of the crystals.
\end{itemize}

Once these parameters have been fixed, then the performance depends on the crystals (material and reflection) and on their orientation accuracy. In this section we focus on the relation between the three above mentioned parameters and the orientation accuracy of the crystals. 
In order to realize a systematic study, one shall have a parameter to optimize. In our case, it is the sensitivity of the resulting telescope. Disregarding the focal plane instrument (i.e. considering it as ideal: pixels infinitely small and a detection efficiency equal to 1), the aim is to put the maximum signal in the smallest area. Doing the assumption that the instrumental background is uniform over the detector plane, the detection significance of a given source is proportional to:
\begin{eqnarray}
n_{\sigma} \propto { A_{eff} \, \epsilon_{fs} \over \sqrt{A_{in}}     }
\end{eqnarray}
with $A_{eff}$ the effective area of the lens, $\epsilon_{fs}$ the fraction of the signal that should be taken into account and $A_{in}$ the area of the focal plane delimiting the fraction $\epsilon_{fs}$ of the signal. $\epsilon_{fs}$ and $A_{in}$ are linked and are calculated in order to maximize the ratio $\epsilon_{fs} / \sqrt{A_{in}}$. 

Since the lens has a rotational symmetry (with the line of sight as axis), the point spread function (PSF) have the same symmetry and consequently $A_{in}$ is a disc. So instead of considering the square root of the area on the focal plane, we can consider the radius of the disc $r_{in}$.

Thus for any lens configuration that we explore, the PSF is computed and the so-called \textit{factor of merit} (FM) is extracted:
\begin{eqnarray}
FM = { A_{eff} \, \epsilon_{fs} \over r_{in}     }.
\end{eqnarray}
On this purpose a semi-analytic code (non Monte-Carlo) has been developed in order to be able to explore a lot of configurations in an acceptable computing time (typically a few seconds per configuration).

To study in a systematic manner the evolution of FM, we have considered always the same objective: to focus in the 300 keV - 400 keV energy band with Cu mosaic crystals using the 111 reflection (with the mean size of the crystallites equal to 50 $\mu$m). Depending on the focal distance and on the crystals size, the code determines the number of rings to be filled with the Cu crystals and their radii. Once the lens is fixed, the PSF is computed. 

\begin{figure}[t]
\begin{center}
\includegraphics[width=0.49\textwidth]{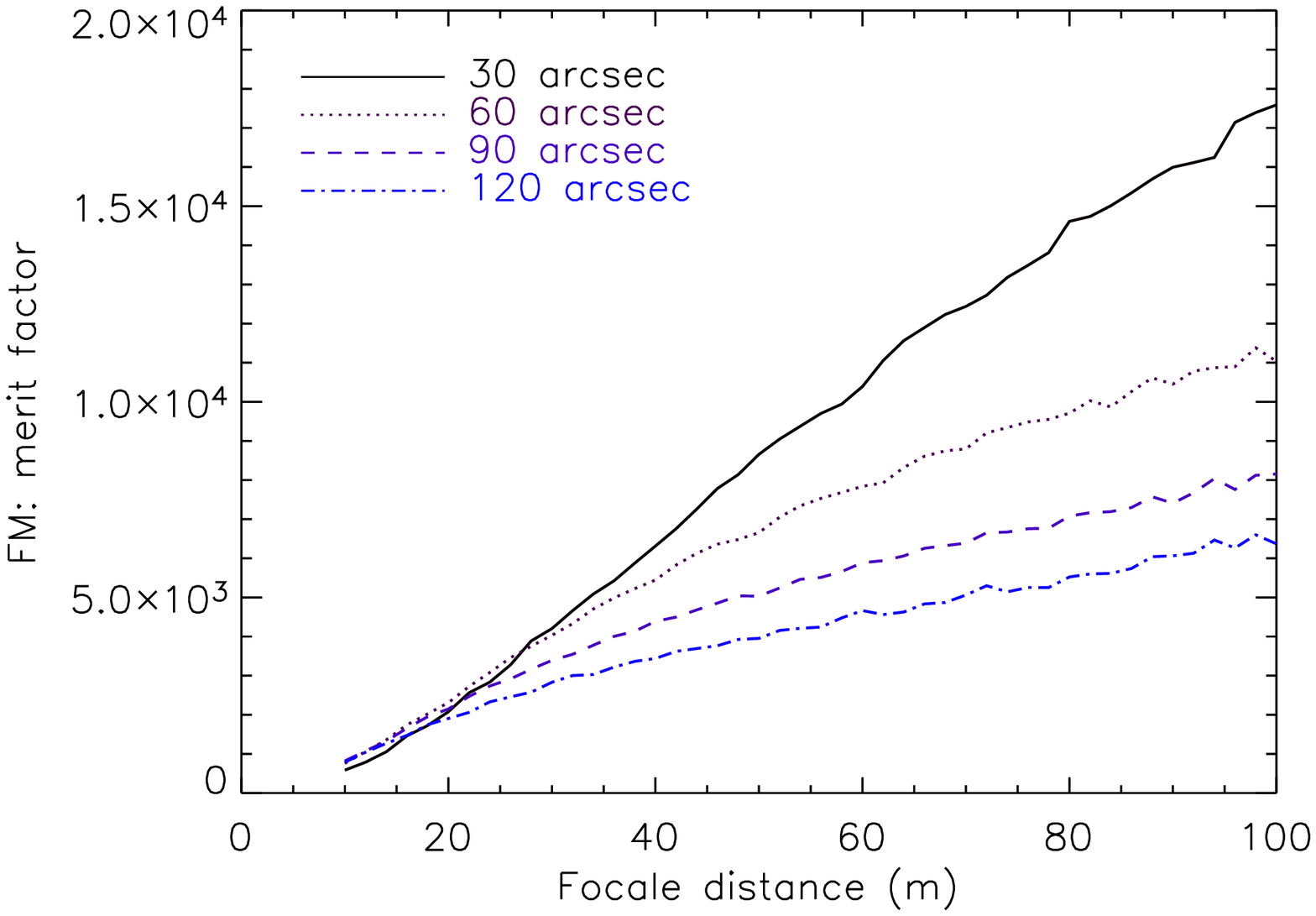}
\includegraphics[width=0.49\textwidth]{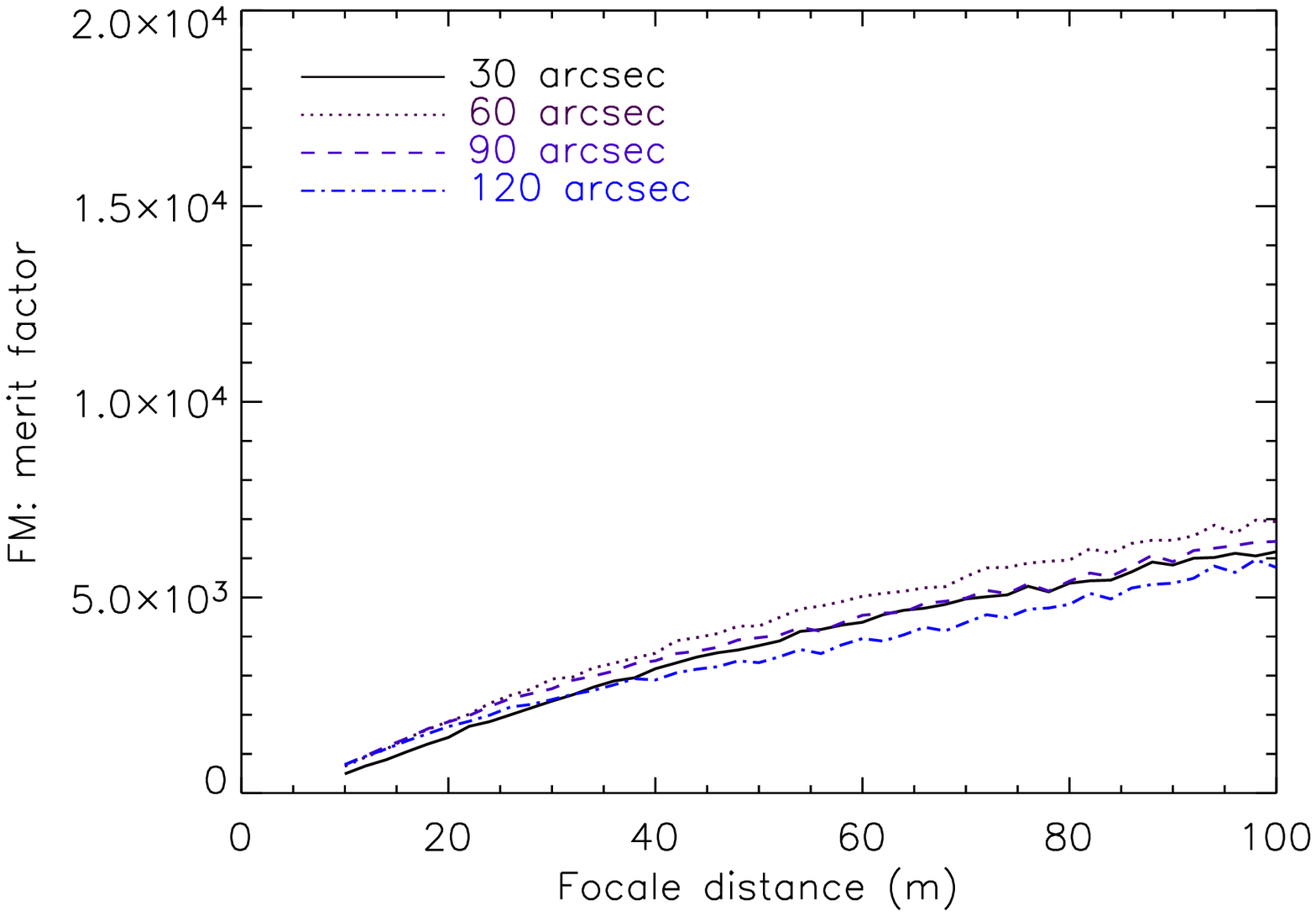}
\caption{Factor of merit versus the lens focal length for  four crystals mosaicity values (30, 60, 90 and 120 arcsec). The left panel shows the case of ideally oriented crystals while the right panel shows degraded case where crystals' orientations are distributed around their nominal angles according to a Gaussian statistics of 30 arcsec of standard deviation.}
\label{fig:MF_vs_focal}
\end{center}
\end{figure}

In the first case, the size of the crystals have been fixed to 15 x 15 mm$^2$, and we investigate the evolution of FM with the focal distance for four mosaicity values. Figure \ref{fig:MF_vs_focal} presents the results in two cases: in the left panel, crystals are ideally orientated while in the right panel their orientation is distributed around the ideal value according to a Gaussian of 30 arcsec of standard deviation ($\sigma$).
Two points are clearly striking: i) in any case the longer focal distance produces the most sensitive telescope. However it has to be balanced with the fact that a longer focal distance requires more crystals  and consequently is more expensive and more heavy. ii) In the ideal case, the smaller mosaicity is the best, but in a more realistic case where crystals are not perfectly oriented, this is no longer true. 
Crystals having a larger mosaicity diffracts more signal (the effective area of the lens grows with the mosaicity of crystals). 
But in the case of a non-ideal lens, the mosaicity of crystals does not influence any more the sensitivity, at least in a range of mosaicity going up to to 3-4 $\sigma$ (in this case, it is probably not a general rule).
Thus better consider the highest mosaicity which does not decrease the sensitivity, since it allows more signal on the focal plane. Moreover crystals having a larger mosaicity are generally easier to produce (not true in the case of CDP crystals though). Consequently it is very important to anticipate properly the orientation precision achievable, in order to determine accordingly the best compromise for the mosaicity.

\begin{figure}[t]
\begin{center}
\includegraphics[width=0.43\textwidth]{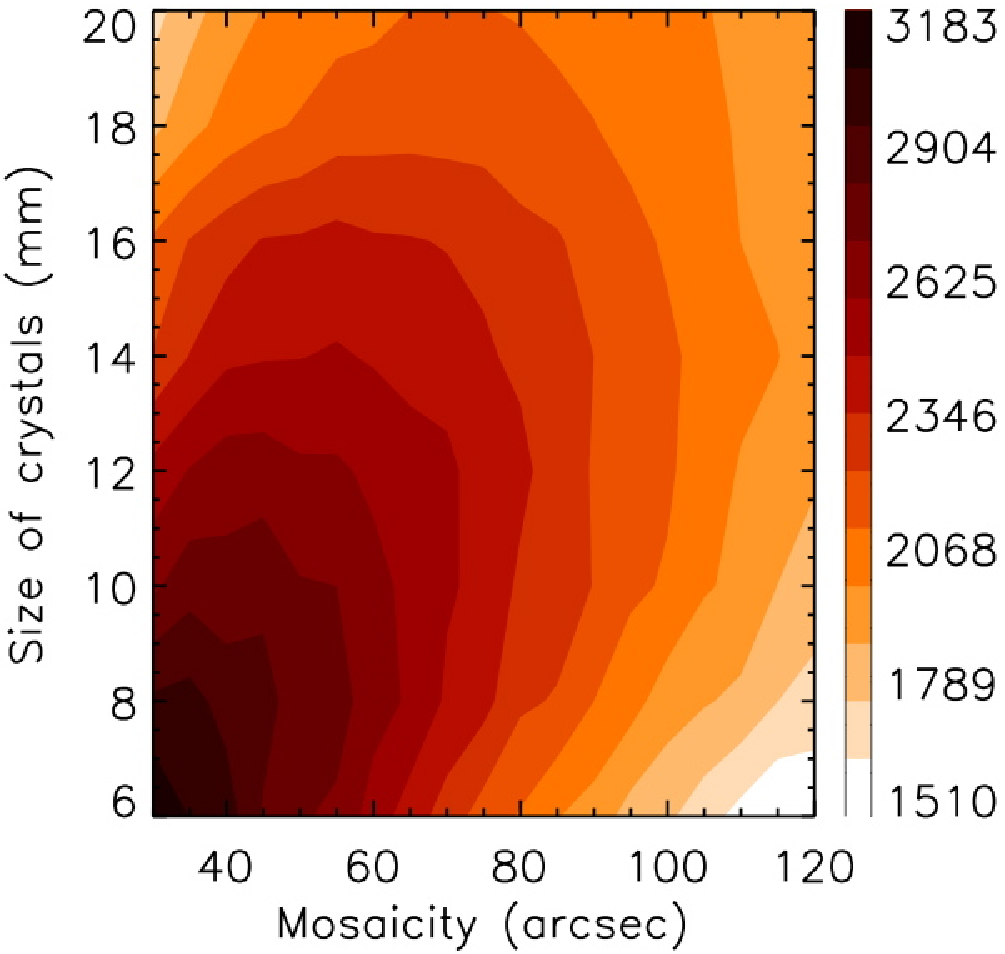}
\hspace{0.5cm}
\includegraphics[width=0.43\textwidth]{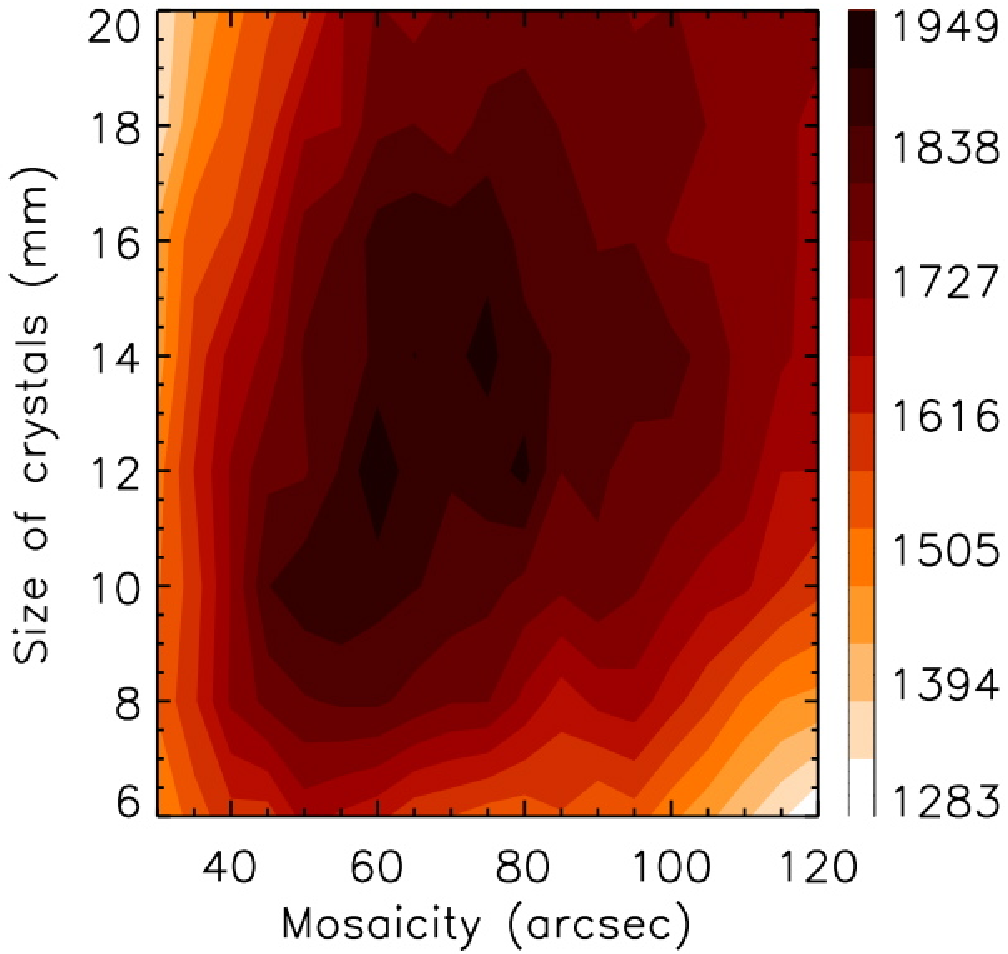}
\caption{Maps of factor of merit as a function of crystals mosaicity and size. The focal length is fixed to 20 m. Like in Figure \ref{fig:MF_vs_focal} the left panel presents the case of crystals ideally orientated while the right panel shows the case of crystals orientation is distributed around their nominal angle according to a gaussian statistics of 30 arcsec of standard deviation.}
\label{fig:MF_vs_Size-Mos_f=20m}
\end{center}
\end{figure}

As a second step, the focal length has been fixed to 20 m (see Sec. \ref{sec:LL20m}) and the influence of crystals mosaicity and size is investigated. Again the both case of ideally orientated crystals and crystals slightly disoriented (Gaussian distribution around the nominal angle, with a standard deviation of 30 arcsec) are considered. Figure \ref{fig:MF_vs_Size-Mos_f=20m} presents the results. In the left panel (ideal case) one sees that the highest FM is yielded by the smallest crystals with the smallest mosaicity. On the other hand in the right panel (degraded case) the maximum factor of merit is much more broad and shifted towards larger mosaicity and size. It is important to notice that both plots (Figure  \ref{fig:MF_vs_Size-Mos_f=20m} left and right) don't have the same scale: The maximum FM in the ideal case is more than 50\% higher than in the degraded case.  This study shows the importance of crystals orientation accuracy for the resulting telescope sensitivity. Two European teams are currently addressing this issue, one in France (see Ref. \citenum{rousselle.2009}) and the other in Italy (see Ref. \citenum{guidorzi.2009ud}). Their goal is to reach a precision of the order of  10 arcsec.

\section{A 20 m FOCAL LENGTH TELESCOPE} 
\label{sec:LL20m}

\subsection{Presentation}

\begin{figure}[t]
\begin{center}
\rotatebox{90}{
\includegraphics[width=0.4\textwidth]{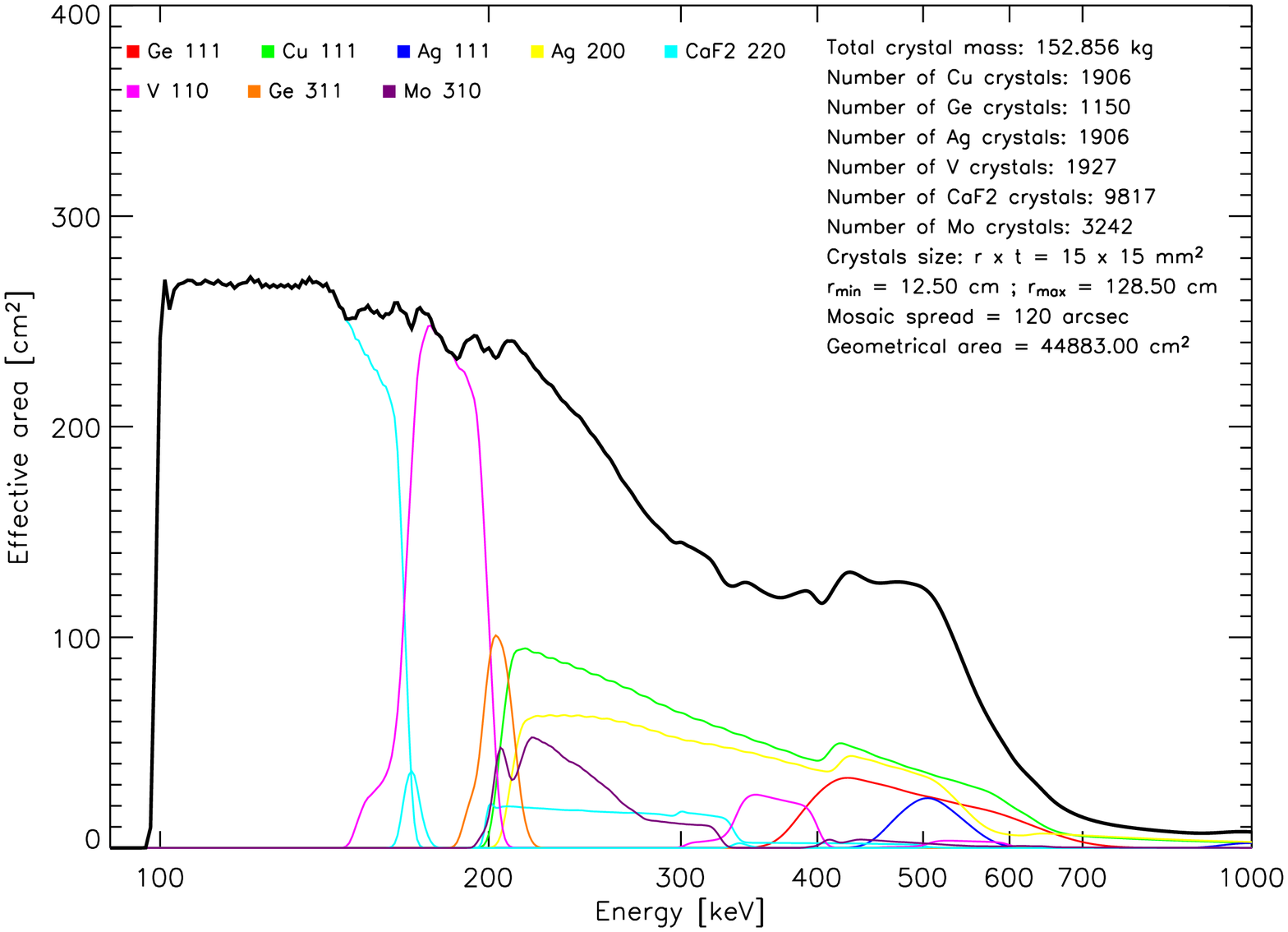}}
\hspace{0.3cm}
\includegraphics[width=0.40\textwidth]{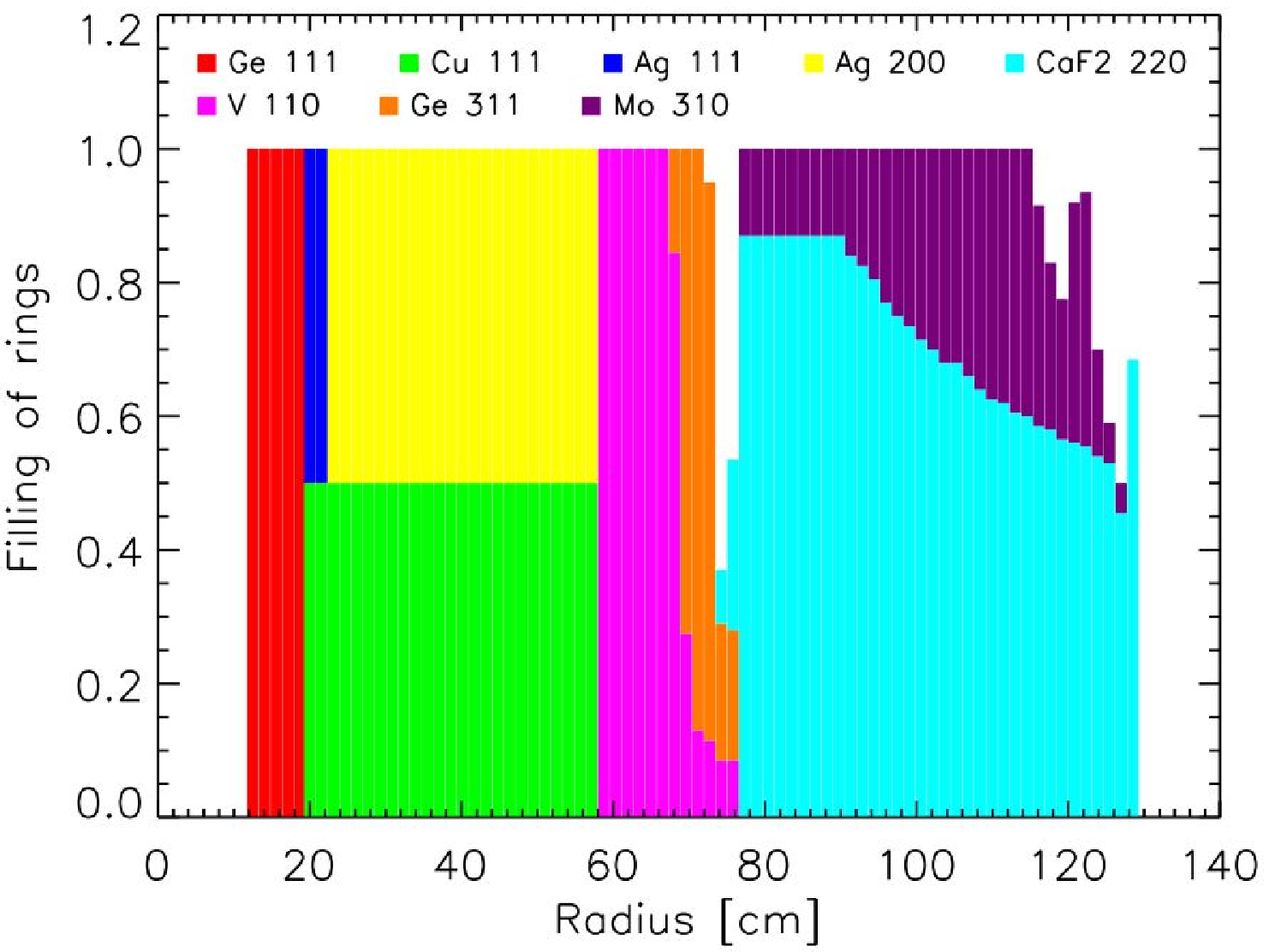}
\caption{\textit{Left.} Effective area of the lens. Colors show the contribution of each material. \textit{Right.} Distribution of the various crystals and reflections over the concentric rings that compose the lens.}
\label{fig:20mLL}
\end{center}
\end{figure}

Would it be possible to build an efficient focusing telescope to study non-thermal emission processes in compact objects, nuclear decay lines in novae and electron - positron (e$^+$ - e$^-$) annihilation without requiring two satellite flying in formation? It is in the idea of answering this question that the lens presented in this section has been designed. As a results, we obtained a lens focusing between 100 keV and 600 keV with a focal length of 20 m, making use of a large variety of crystals in order to yield the best reflectivity in each energy sub-range. The lens is composed of about $20\,000$ crystal tiles of 15 x 15 mm$^2$ each and of thickness calculated to maximize the reflectivity (depending on the crystalline material, the reflection and the energy it diffracts). Crystals selected for this lens are Ge, Cu, Ag, CaF$_2$, V, and Mo. They  are all mosaic crystals with a mosaicity of 2 arcmin. Figure \ref{fig:20mLL} shows the distribution of the crystals over the rings and the effective area it produces. The contribution to the total effective area of each material and reflection is plotted with dedicated colors. 
The effective area equals 270 cm$^2$ at 100 keV and decreases down to $\sim$ 130 cm$^2$ at 500 keV to finally fall under 45 cm$^2$ beyond 600 keV (which is not negligible, considered the area of the focal spot, see next section).

Table \ref{tab:LLparameters} gives numerical parameters of the lens design. The crystals cover a disc of about 4.5 m$^2$ area. Based on the former studies of mission featuring a Laue lens (MAX\cite{duchon.2005yf} and GRI\cite{knodlseder.2009kx}), the weight of the lens structure can be evaluated to $\sim$ 70 kg/m$^2$. It results that our lens, crystals and carrying structure included would weigh $\sim$ 470 kg. The focal distance could be achieved by an extensible boom, which allows the telescope to fit in the envelop of a medium size mission.

\begin{table}[h]
\begin{center}
\begin{tabular}{ll}
\hline
Focal distance (m) & 20 \\
Mosaicity of crystals (arcmin) & 2 \\
Total number of crystals & $20\,000$ \\
Total crystal mass (kg) & 150 \\
Inner radius (m) & 0.125 \\
Outer radius (m) & 1.285 \\
Collecting area (m$^2$) & 4.5 \\
\hline 
\\
\end{tabular}
\caption{Parameters of the 20 m focal length Laue lens. }
\end{center}
\label{tab:LLparameters}
\end{table}%

Crystals used in this design were studied theoretically\cite{barriere.2009sf} , but not all of them have yet proved experimentally to be obtainable with the proper quality: V and Mo have not been measured and we measured only one sample of CaF$_2$ so far. But the interest of this design is that its performance is robust: 120 arcsec of bandpass is not a tight constraint for mosaic crystal. But more importantly with 15 x 15 mm$^2$ crystals and a mosaicity of 120 arcsec, the lens performance won't degrade if an orientation accuracy of $\sigma$ $\leq$ 30 arcsec (Gaussian distribution) can be achieved, which does not seem out of reach (See Ref. \citenum{guidorzi.2009ud} in this session).

\subsection{Performance estimates}
The sensitivity estimate is calculated considering a 10 x 10 cm$^2$ focal plane made of 3 stacked pixellated CdTe or CdZnTe layers. Pixels size are 1.6 x 1.6 mm$^2$, and each layer measures 20 mm thick and is segmented in four pseudo-layers. This focal plane instrument is directly inspired from the one that was proposed for the GRI, at the difference that in the present case it is smaller (see Ref. \citenum{natalucci.2008kl}). Background level has been scaled down by a factor of three from INTEGRAL/ISGRI measured instrumental background to take into account a low Earth orbit and a smaller satellite (L. Natalucci, private communication). 

\begin{figure}[t]
\begin{center}
\rotatebox{90}{
\includegraphics[width=0.4\textwidth]{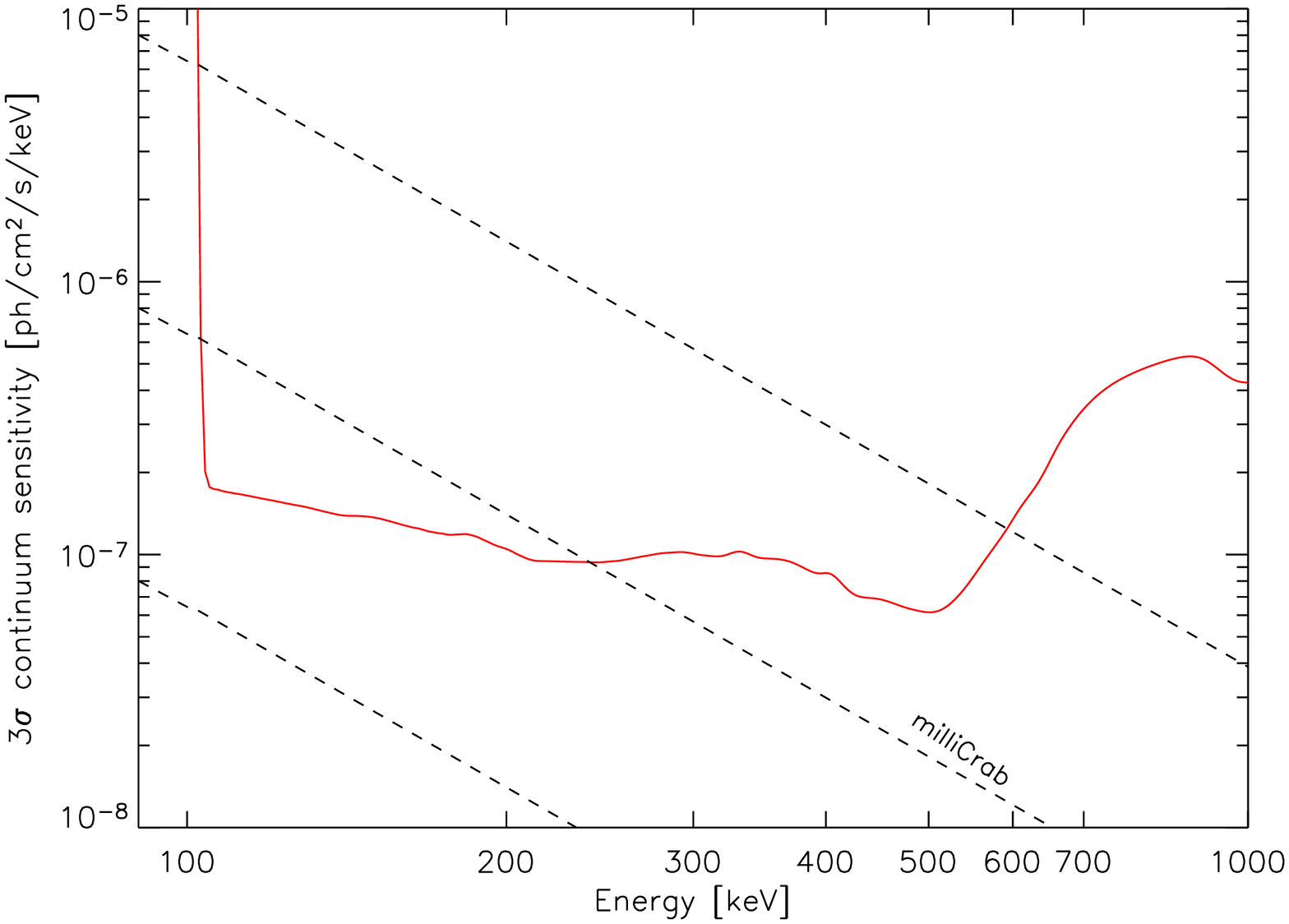}}
\hspace{0.5cm}
\includegraphics[width=0.35\textwidth]{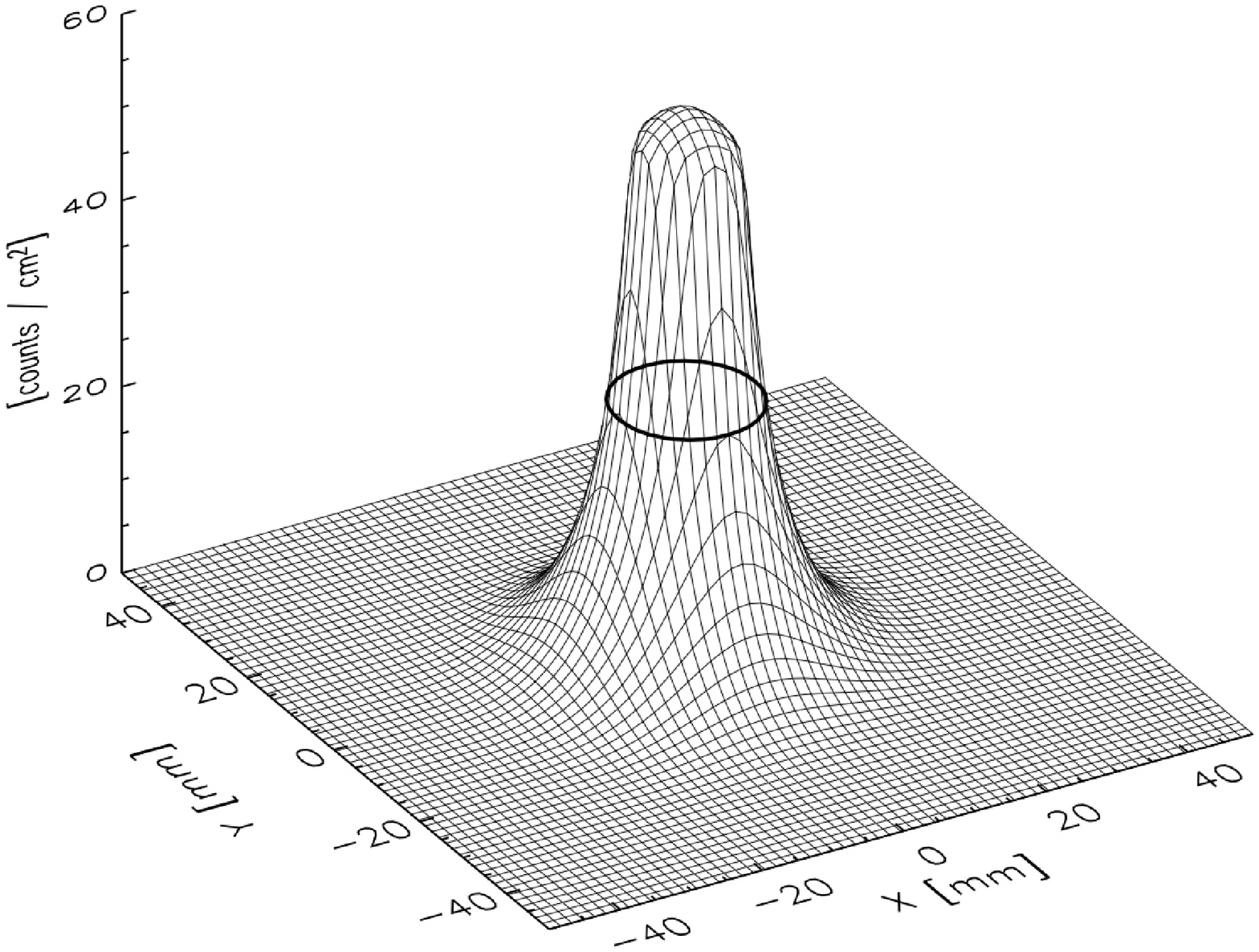}
\caption{
\textit{Left.} 3$\sigma$ continuum sensitivity ($\Delta E = E / 2$) of the telescope for 100 ks observation time. See text for details.  
\textit{Right.} PSF of the lens for a point source on-axis. $r_{in}$ is equal to 10.6 mm and encircles $\epsilon_{fs}$ = 59\% of the total signal (see Sec. \ref{sec:design} for the definitions of $r_{in}$ and $\epsilon_{fs}$).
}
\label{fig:LL20sensitivity}
\end{center}
\end{figure}

Figure \ref{fig:LL20sensitivity} shows the resulting 3$\sigma$ sensitivity for a continuum ($\Delta E = E/2$) in 100 ks exposure time. It starts from 2 x 10$^{-7}$ ph/s/cm$^2$/keV at 100 keV and reach 6 x 10$^{-8}$ ph/s/cm$^2$/keV at 500 keV.

Looking at the PSF (Fig. \ref{fig:LL20sensitivity}, right panel) and the size of the focal plane instrument, one can expect a field of view of 13 arcmin (FWHM) and an angular resolution of $\sim$ 3.5 arcmin. This quite coarse angular resolution with respect to the one foreseen for the GRI mission is due to the fact that in the present case the ratio crystal size to focal distance is five times higher. However, in this energy range this performance would constitute a sizable step forward with respect to currently operating missions. We shall precise that the performance of Laue lens telescopes could even be enhanced if a Compton telescope is put at the focus\cite{weidenspointner.2005rw}. It this case it has been shown that the background rejection is improved leading to a sensitivity increased by a factor of two. On the other hand, if the first layers of the focal plane Compton camera are made of Silicon, it allows polarization measurement down to $\sim$ 100 keV.

A simulation code is currently being developed to deepen our understanding of the imaging capabilities of Laue lenses. A quite simple version based on maximum likelihood analysis of the focal plane is giving its first results. So far it permits to find easily a point source in the field of view (blind search, no information on the source position or flux is provided), and to reconstruct the image of an extended emission if several pointings are combined. Future work includes the development of an Iterative Removal of Source (IROS) algorithm to handle several point sources in a single pointing and a proper background removal method.

\begin{figure}[t]
\begin{center}
\includegraphics[width=0.32\textwidth]{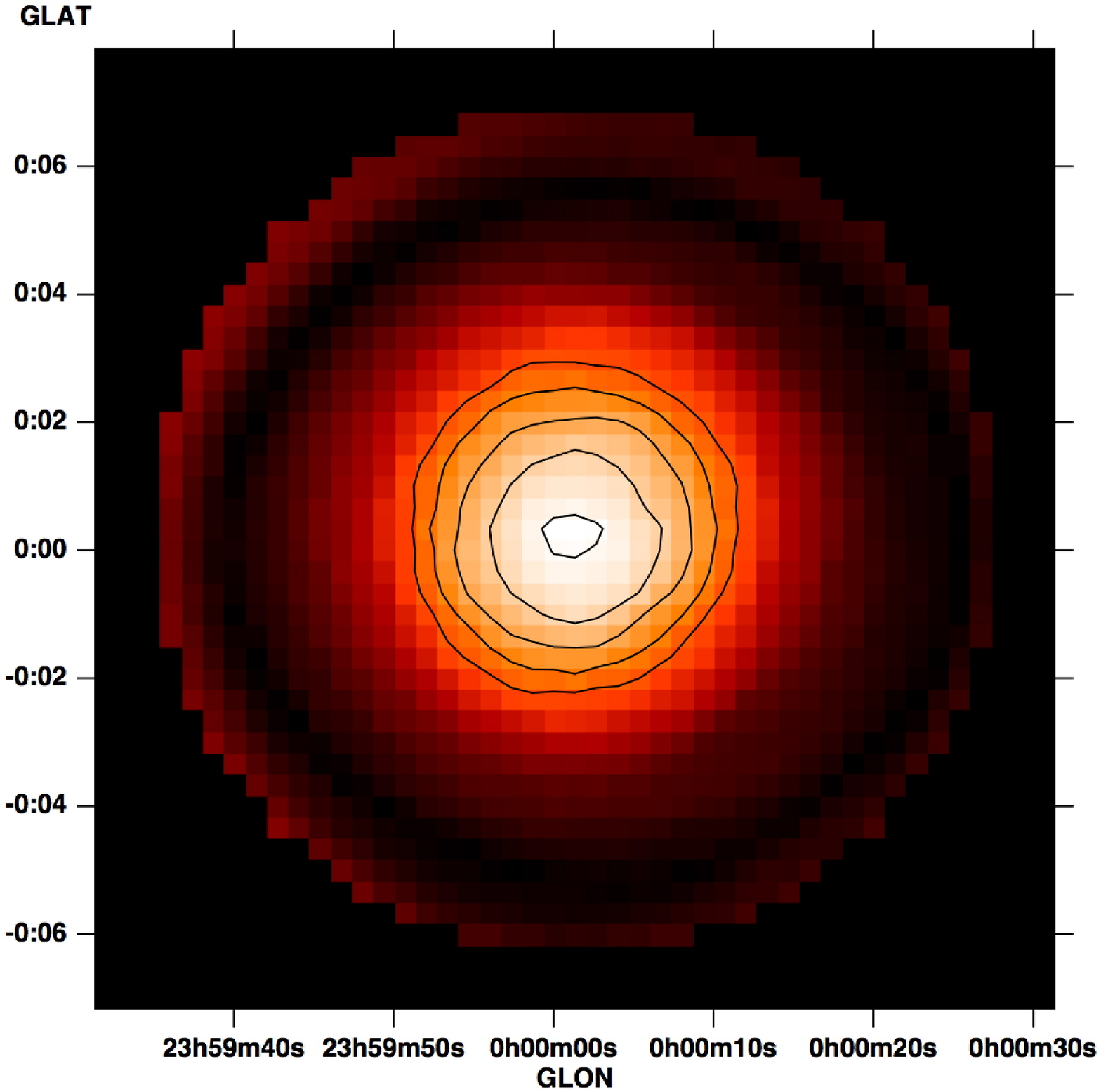}
\includegraphics[width=0.32\textwidth]{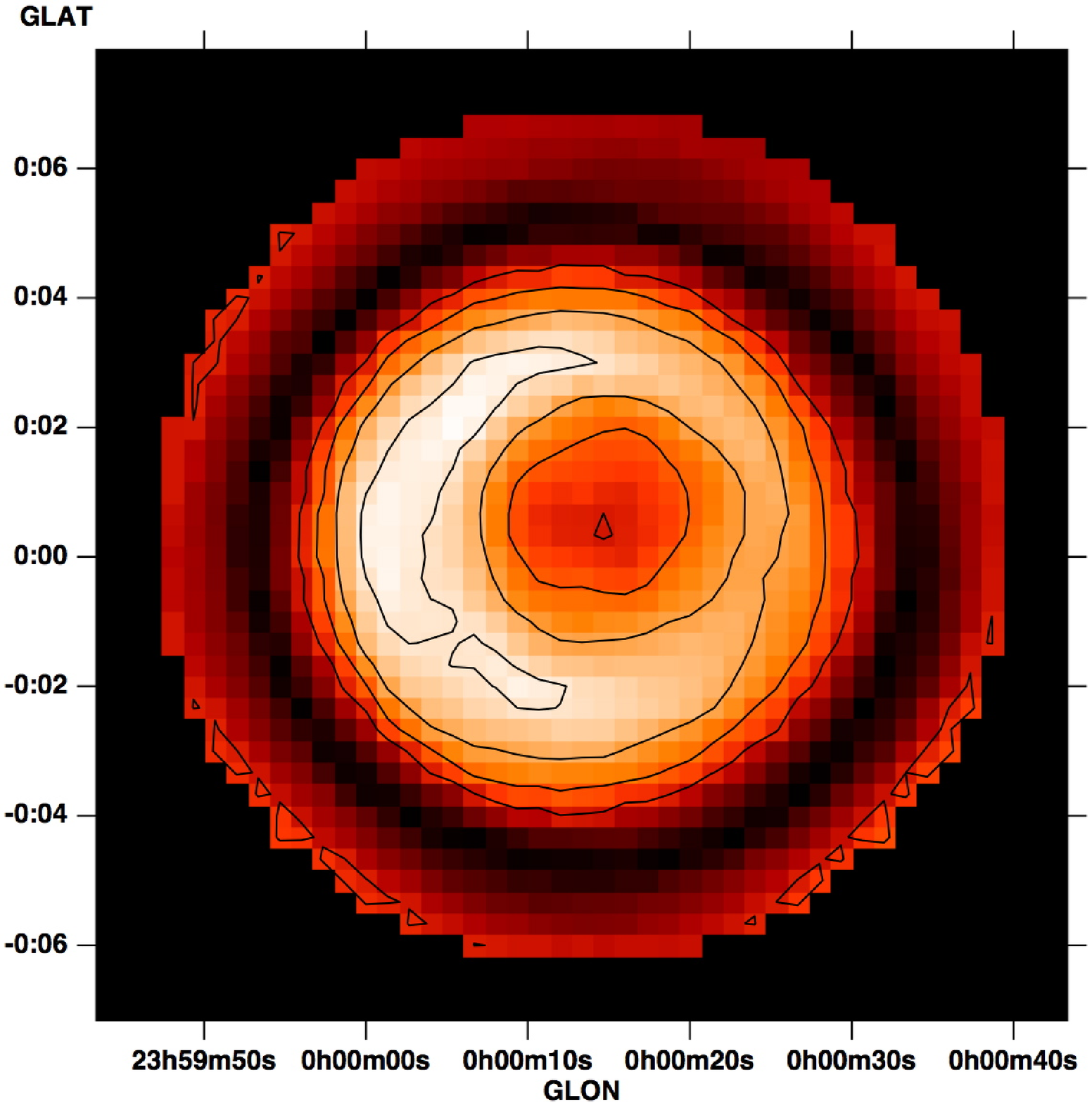}
\includegraphics[width=0.32\textwidth]{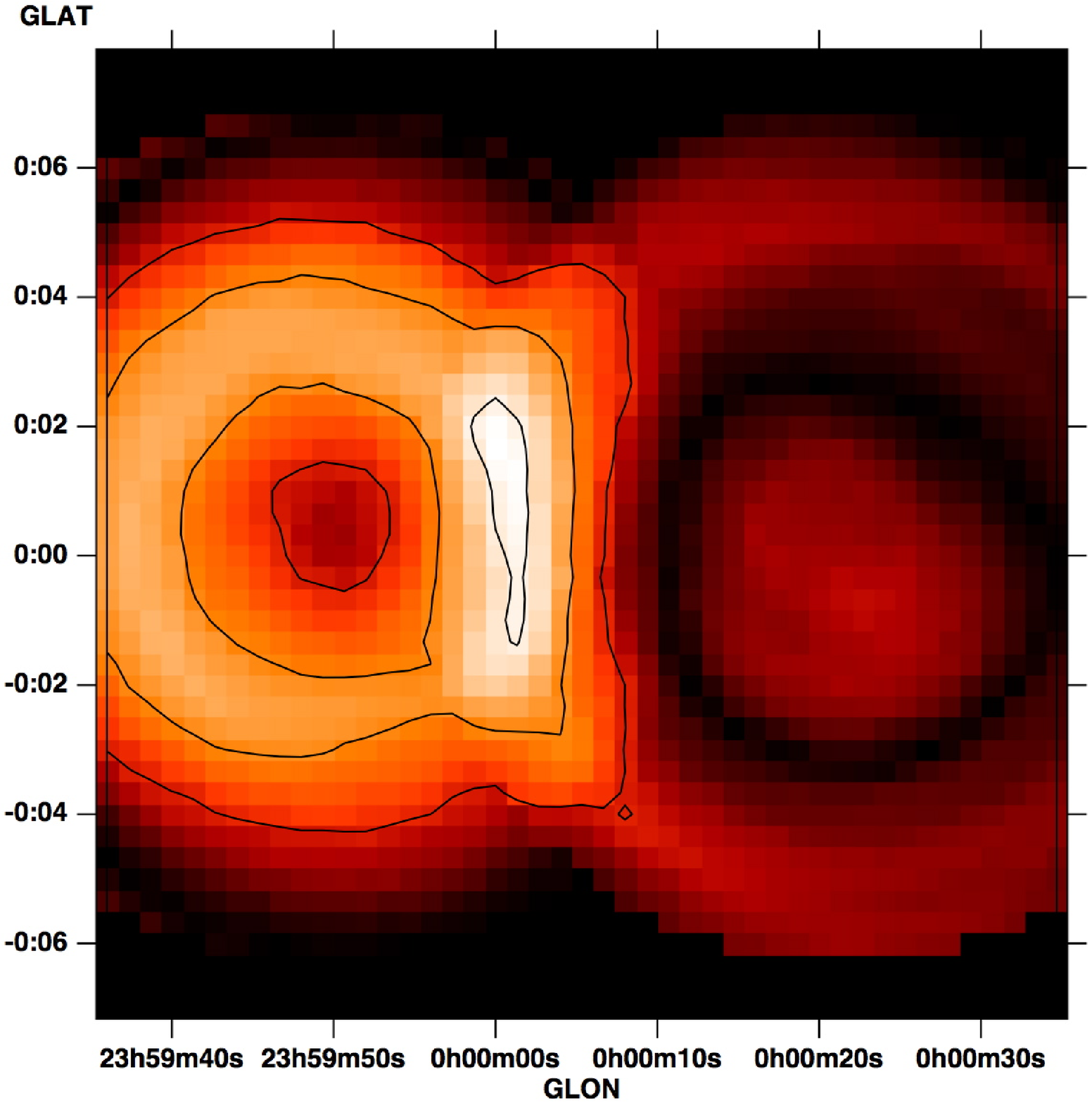}
\caption{Detection significance of a point source having a flux of 1 mCrab (placed aarbitrarily at the Galactic center). Each map extends over 20 x 20 arcmin and each bin measures 20 x 20 arcsec. Contours show step of 2 $\sigma$. See text for details}
\label{fig:pt_src_obs}
\end{center}
\end{figure}

Figure \ref{fig:pt_src_obs} presents the case of the observation during 100 ks in the 100 keV - 150 keV range of a single point source that emits a flux of 1 mCrab. Instrumental background on the focal plane is kept as defined previously in this section. The left panel of Figure \ref{fig:pt_src_obs} shows the case where the source is lying on the line of sight of the telescope. In this case it is very well detected with of confidence level of 18 $\sigma$ (contours shows step of 2 $\sigma$), and quite well located within $\sim$ 1 arcmin. The central panel shows the case where the source is placed 3 arcmin off-axis. Here a rotational degeneracy appears in its positionning, but the source is still detected with a level of confidence of 11 $\sigma$. The last case (right panel of Fig.  \ref{fig:pt_src_obs}) illustrates that the degeneracy can be limited by a dedicated observation strategy. In this case, the exposure time has been shared in two pointings directed 3 arcmin and 5 arcmin far from the source. Detection significance falls to 7.5 $\sigma$ but the positioning of the source is improved.

Finally, the observation of an extended region emitting in the e$^+$ - e$^-$ line at 511 keV is simulated. We considered a dummy source having a bi-dimensional Gaussian shape (standard deviation of 2 arcmin x 7 arcmin) inclined of 45$^{\circ}$ from the plane of the instrument, with a total integrated flux of 1 x 10$^{-4}$ ph/s/cm$^2$ in a 2 keV wide line. A spectral resolution of 2\% and a background level of 1.3 x 10$^{-4}$ cts/s/cm$^3$/keV  are assumed for the focal plane detector at 511 keV. It results a background count rate of 1.3 x 10$^{-3}$ cts/s/cm$^{3}$ in the line, uniformly distributed (Poisson statistics) over the pixels of the focal plane. 

\begin{figure}[htbp]
\begin{center}
\includegraphics[width=0.45\textwidth]{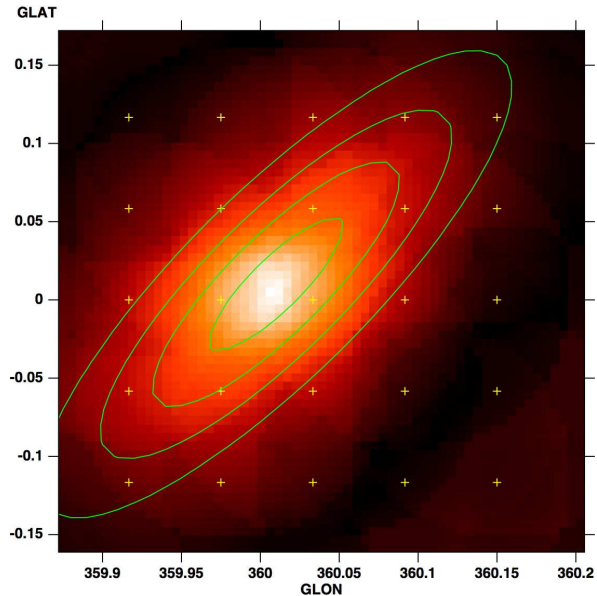}
\caption{Detection significance of a (dummy) extended source having a total flux of 1 x 10$^{-4}$ ph/s/cm$^2$ in a narrow line centered on 511 keV. The contours show the original source shape, and the crosses shows the pointings composing the observation pattern. The sky map extends over 20 x 20 arcmin , and each bin measures 20 x 20 arcsec.}
\label{fig:ext_src_obs}
\end{center}
\end{figure}

To resolve the structure of the emission, it is necessary to divide the observation time in many pointings to break the degeneracy due to the small asymmetry of the lens response for off-axis sources. Figure \ref{fig:ext_src_obs} shows the reconstructed image of the sky, the dithering pattern that has been used for the observation (for a total exposure time of 500 ks) and the contours of the original dummy extended source. Despite the strong background the shape of the source is quite well reconstructed. The center of the source is detected with a significance of 9.5 $\sigma$. This study demonstrates the imaging capabilities of Laue lenses, which can cover objects of a few arcmin of spatial extent with an angular resolution of the order of 3 arcmin in the case of the lens presented in this paper.

\section{CONCLUSION} 
\label{sec:conclusion}

Thanks to the Italian and French endeavor, Laue lenses are becoming a reality. The activities for the development of efficient crystals has already proven that Cu and SiGe crystals could be produced with the proper quality and bandpass range (mosaicity). Our latter experiments at ESRF and ILL showed that other materials like Ag, Au, Rh and GaAs are very good candidates to enlarge our portfolio of qualified crystals. Recently, the technique of grooved crystals opened a new path to realize CDP crystals with very promising results. 

In the meantime the study of Laue lenses design make also progresses. We have used these knowledge to design a \textit{robust} lens focusing in the 100 keV - 600 keV range. Despite the fact it is based on crystals not all proved yet, the size of crystals, their mosaicity, and the focal length of 20 m make this lens does not seem out of reach considering the currently available technology. One important point is that the performance presented in this paper - a sensitivity better than 2 x 10$^{-7}$ ph/s/cm$^2$/keV for 100 ks exposure - is not sensitive to the orientation precision of crystals, provided it stays within a gaussian distribution of $\sigma$ = 30 arcsec. This constraint is much relaxed with respect to the goals of the two current teams building Laue lens prototypes (Ref. \citenum{rousselle.2009} and \citenum{guidorzi.2009ud}, this session).

The recent investigations on imaging capabilities confirm that it is possible to perform high angular resolution imaging with Laue lenses, despite they are not direct imaging optics.

\section*{ACKNOWLEDGMENTS} 
NB and LN are grateful to ASI for the support of Laue lens studies through grant I/088/06/0.


\bibliography{Barriere_SPIE09}   
\bibliographystyle{spiebib}   

\end{document}